\documentstyle[sprocl]{article}

\begin{document}
\baselineskip=.285in

\title{Conformal Phase Transition}
\author{
{\sc V. A. Miransky}\\
{\it Institute for Theoretical Physics}\\
{\it 252143 Kiev, Ukraine}\\
 }

\maketitle

\begin{abstract}
The conception of the conformal phase transiton (CPT), which is
relevant for the description of non-perturbative dynamics in
gauge theories, is discussed.
\end{abstract}

\section{Introduction}

In this talk I will discuss the conception of the conformal phase
transition (CPT) which has been recently introduced and elaborated
in a work done together with Koichi Yamawaki \cite{0}.As I will try
to convince you, this conception is relevant for the description of
non-perturbative dynamics in gauge theories.

The standard framework for the description of continuous phase transitions
is the Landau-Ginzburg, or $\sigma$-model-like, effective action \cite{1}.
In particular, in that approach, a phase transition is governed
by the parameter
\begin{equation}
M^{(2)} \equiv \frac{d^2 V}{d X^2} \left. \right|_{X = 0} \; , \label{1}
\end{equation}
where $V$ is the effective potential and $X$ in an order parameter
connected with the phase transition.
When $M^{(2)} > 0 \quad (M^{(2)} <0) $, the symmetric (non-symmetric) phase
is realized.
The value $M^{(2)} = 0$ defines the critical point.

Thus, as $M^{(2)}$ changes, one phase smoothly transforms into another.
In particular, masses of light excitations are continuous
(though non-analytic at the critical point) functions of
such parameters as coupling constants, temperature, etc.

If $M^{(2)} \equiv 0$, the parameter
$M^{(4)} \equiv \frac{d^4 V}{d X^4} \left. \right|_{X = 0}$
plays the role of $M^{(2)}$, etc.

In this talk, I will describe a non-$\sigma$-model-like,
though continuous, phase transition, which is relevant
for the description of non-perturbative dynamics in gauge field theories.
Because, as will become clear below, this phase transition is intimately
connected with a nonperturbative breakdown of the conformal symmetry,
we will call it
the conformal phase transition (CPT).

In a $\sigma$-model-like phase transition, around the crirical point
$z=z_c$ (where $z$ is a generic notation for parameters of a theory,
as the coupling constant $\alpha$, number of particle flavors $N_f$, etc),
an order parameter $X$ is
\begin{equation}
X = \Lambda f(z) \label{2}
\end{equation}
($\Lambda$ is an ultraviolet cutoff),
where $f(z)$ has such a non-essential singularity at $z=z_c$
that $\lim f(z)=0$ as $z$ goes to $z_c$ both in symmetric and
non-symmetric phases. The standard form for $f(z)$ is
$f(z) \sim (z - z_c)^{\nu}$, $\nu > 0$, around $z = z_c$.
\footnote {
Strictly speaking, Landau and Ginzburg considered the mean-field
phase transition with $\nu = 1/2$.
By a $\sigma$-model like phase transition, we
understand a more general class, when fields may have anomalous
dimensions \cite{3}.
}

The CPT is a very different continuous phase transition. We define it
as a phase transition in which an order parameter $X$ is given by
Eq. (\ref{2})
where $f(z)$ has such an $\underline{\mbox{essential}}$ singularity
at $z=z_c$ that while
\begin{equation}
\lim_{z \to z_c} f(z) =0 \label{3}
\end{equation}
as $z$ goes to $z_c$ from the side of the non-symmetric phase,
$\lim f(z) \ne 0 $
as $z \to z_c$ from the side of the symmetric phase
(where $X \equiv 0$).
Notice that since the relation (\ref{3}) ensures that
the order parameter $X \to 0$ as $z \to z_c$,
the phase transition is continuous.

There actually exist well-known models in which such a phase
transition is realized.
As an example of the CPT is the phase transition at $\alpha^{(0)} =0 $
($\alpha^{(0)} = \frac{(g^{(0)})^2}{4\pi}$ is the bare coupling constant)
in massless QCD with a small, say, $N_f \leq 3$, number of fermion flavors.
In this case, the order parameter $X$, describing chiral symmetry breaking,
is $X \sim \Lambda_{\mbox{QCD}}$ and
\begin{equation}
X \sim \Lambda_{\mbox{QCD}} \sim \Lambda f(\alpha^{(0)}) \; , \label{4}
\end{equation}
where $ f(\alpha^{(0)}) \simeq \exp \left( -\frac{1}{b \alpha^{(0)}} \right)$
($b$ is the first coefficient of the QCD $\beta$ function).
The function $ f(\alpha^{(0)}) $ goes to zero only if $\alpha^{(0)} \to 0$
from the side of $Re \alpha^{(0)} > 0$.

The above example is somewhat degenerate:
the critical point $\alpha^{(0)}_c =0$ is at the edge of the physical space
with $ \alpha^{(0)} \geq 0$.
A more regular example of the CPT is given by the phase transition at
$g^{(0)} = 0 $ in the $(1+1)$-dimensional Gross-Neveu model:
in that case both positive and negative values of $g^{(0)}$ are physical
(see Sec.3).

There may exist more sophisticated realizations of the CPT.
As is discussed in \cite{0}, an example of the CPT may be provided
by the phase transition with respect to the number of fermion flavors $N_f$
in a $SU(N_c)$ vector-like gauge theory in $(3+1)$ dimensions, considered
by Banks and Zaks long ago \cite{2}.
In that case, unlike the phase transition at $\alpha^{(0)}=0$ in QCD,
the critical value $N_f^{cr}$ separates two physical phases,
with $N_f < N_f^{cr}$ and $N_f \geq N_f^{cr}$.

There may exist other examples of the CPT.
Also there may exist phase transitions in $(2+1)$-dimensional theories
which ``imitate'' the CPT (see Sec.5).

The main goal of this talk is to reveal the main features of the
CPT (common for its different realizations).

The CPT is $\underline{\mbox{not}}$
a $\sigma$-model-like phase transition, though it is continuous.
In particular, in the CPT, one cannot introduce the parameters
$M^{(2n)} = \frac{d^{2n} V}{d X^{2n}} \left. \right|_{X=0} \; ,
n=1,2, \cdot \cdot \cdot \; , $ governing the phase transition.
Another characteristic feature of the CPT is an abrupt
change of the number of
light excitations as the critical point is crossed
(though the phase transition is continuous). While evident in QCD
and the Gross-Neveu model, it is realized in a more subtle way
in the general case.
This feature implies a specific form of the effective
action describing
light excitations in theories with the CPT, which will be discussed
in this talk later.

\section{Peculiarities of the Spectrum of Light Excitations in the CPT}

As was already pointed out in the Introduction, in the
case of the $\sigma$-model-like
phase transition, masses of light excitations are continuous functions
of the parameters $z$ around the critical point $z=z_c$
(though they are non-analytic at $z=z_c$).
Let us show that the situation in the case of the CPT is different:
there is an abrupt change of the spectrum of light excitations,
as the critical point $z=z_c$ is crossed.

Let us start from a particular, and important, case of the CPT connected
with dynamical chiral symmetry breaking. In this case, in the non-symmetric
phase, amongst light (with masses much less than cutoff $\Lambda$)
excitations, there are massless Nambu-Goldstone (NG) bosons $\pi$,
their chiral partners, $\sigma$ bosons, and light (with $m_{dyn} \ll \Lambda$)
fermions. The masses of $\sigma$ and fermions are given by scaling relations:
\begin{eqnarray}
M_{\sigma}^2 &=& C_{\sigma} \Lambda^2 f(z) \label{5} \\
m_{dyn}^2 &=& C_{f} \Lambda^2 f(z) \; , \label{6}
\end{eqnarray}
where $C_{\sigma}$ and $C_f$ are some positive constants, and $f(z)$ is a
universal scaling function.
Because of the assumption (\ref{3}), $M_{\sigma}^2$ and
$m_{dyn}^2$ are
indeed much less than $\Lambda^2$, when $z$ is near $z_c$ from the
side of the non-symmetric phase.

Now, are there light $\pi$ and $\sigma$ resonances in the symmetric phase,
with $m_{dyn}=0$?
Since, as was assumed, $\lim f(z) \ne 0 $ as $z \to z_c$ in that phase, one
should expect that there are no light resonances.
Let us show that this is indeed the case.

One might think that in the symmetric phase the mass relation for
$\pi$ and $\sigma$ is yielded by the analytic continuation of the relation
(\ref{5}) for $M_{\sigma}^2$.
However this is not the case. The point is that while in
the non-symmetric phase, $\pi$ and $\sigma$ bosons are described by
Bethe-Salpeter (BS) equations with a non-zero fermion mass, in the symmetric
phase they are described by BS equations with $m_{dyn} \equiv 0$.
Because of that, BS equations (and, more generally, all the Schwinger-Dyson
equations for Green's functions) in the symmetric phase are not yielded
by an analytic continuation of the equations in the non-symmetric phase.

To overcome this obstacles, we shall use the following trick.
In the non-symmetric phase, besides the stable solution with
$m_{dyn} \ne 0$, there is also an unstable solution with $m_{dyn} = 0$.
In that solution, $\pi$ and $\sigma$ bosons are tachyons:
$M_{\pi}^2=M_{\sigma}^2 \equiv M_{tch}^2<0$.
Since the replacement of $m_{dyn} \ne 0 $ by $m_{dyn}=0$
(at fixed values of the parameters $z$) does not change the ultraviolet
properties of the theory, the scaling relation for the tachyon masses has
the same form as that in Eqs. (\ref{5}) and (\ref{6}):
\begin{equation}
M_{\pi}^2=M_{\sigma}^2 = M_{tch}^2 =
-C_{tch} \Lambda^2 f(z), \quad C_{tch}>0.
\label{7}
\end{equation}
Since now $m_{dyn} = 0$, the BS equations for tachyons have the same form
as the BS equations for $\pi$ and $\sigma$ in the symmetric phase;
the difference between these equations is only in the values of $z$
(for convenience, we shall assume that $z > z_c \quad (z < z_c)$ in
the non-symmetric (symmetric) phase).
Then, in the symmetric phase,
\begin{equation}
M_{\pi}^2=M_{\sigma}^2 = -C_{tch} \Lambda^2 f(z) \; ; C_{tch}>0,
\label{8}
\end{equation}
with $z<z_c$ and $C_{tch}$ from Eq. (\ref{7}).
Notice that because in the symmetric phase $\pi$ and $\sigma$ bosons
decay to massless fermions and antifermions, $M_{\pi}^2$ and $M_{\sigma}^2$
are complex, i.e. $\pi$ and $\sigma$ are now resonances,
if they exist at all.

Since, by definition, in the CPT, $\lim f(z) \ne 0 $ as $z \to z_c -0 $,
we conclude from Eq. (\ref{8}) that there are no light resonances near
the critical point from the side of the symmetric phase:
$
| M_{\pi}^2 |= |M_{\sigma}^2| \sim \Lambda^2
$ as $z \to z_c -0 $.

So far, for concreteness, we have considered the case of
dynamical chiral symmetry breaking. But it is clear that
(with minor modifications) this consideration can be extended to
the general case of the CPT connected with spontaneous breakdown of
other symmetries.

Notice also that the relation in Eq. (\ref{8}) can be useful
for general phase
transitions and not just for the CPT.
The point is that the scaling function $f(z)$ can be determined from
the gap equation for the order parameter ($m_{dyn}^2$, in the case of
chiral symmetry) which is usually much simpler than the BS equation for
massive composites. For example,
an abrupt change of the spectrum at the critical point $z=z_c$ have been
revealed in some models: in quenched QED4 \cite{4,5} and QED3 \cite{6}.
This conclusion was based on an analysis of the effective action
\cite{4} and the BS equation \cite{5,6}, considered in a rather crude
approximation. On the other hand, since the determination of the
scaling function $f(z)$ in these models is a much simpler task, this
conclusion can be firmly established in the present approach (see
Secs.4 and 5).
Thus the present consideration yields a simple and general criterion of
such a peculiar behavior of the spectrum of light excitations.

It is clear that the abrupt change of the spectrum discussed above
implies rather peculiar properties of the effective action for
light excitations at the critical point. Below we shall consider
this problem in more detail. We shall also reveal
an intimate connection between this point and the essential difference
of the character of the breakdown of the conformal symmetry
in different phases of theories with the CPT.

\section{$D$-dimensional Nambu-Jona-Lasinio model.
The CPT at D=2}

In this section we consider the dynamics in the $D$-dimensional
$(2 \leq D < 4)$ Nambu-Jona-Lasinio (Gross-Neveu) model and, in particular,
describe the CPT in the Gross-Neveu (GN) model at $D=2$.
This will allow to illustrate main features of the CPT in a very clear way.

The Lagrangian density of the $D$-dimensional GN model, with the
$U(1)_L \times U(1)_R$ chiral symmetry, is
\begin{equation}
{\cal L}= \frac{1}{2} \left[ \bar{\psi}, (i \gamma^{\mu}\partial_{\mu}) \psi
\right]
+ \frac{G}{2} \left[ (\bar{\psi} \psi)^2 + (\bar{\psi} i \gamma_5 \psi)^2
\right] , \label{9}
\end{equation}
where $\mu=0,1, \cdot \cdot ,D-1$, and the fermion field carries
an additional ``color'' index $\alpha = 1,2, \cdot \cdot , N_c$.
The theory is equivalent to the theory with the Lagrangian density
\begin{equation}
{\cal L}' = \frac{1}{2} \left[ \bar{\psi}, (i\gamma^{\mu}\partial_{\mu})
\psi \right] -\bar{\psi} (\sigma + i \gamma_5 \pi) \psi
- \frac{1}{2G} (\sigma^2 + \pi^2) . \label{10}
\end{equation}
The Euler-Lagrange equations for the auxiliary fields $\sigma $ and $\pi$
take the form of constraints:
\begin{equation}
\sigma = -G \bar{\psi}\psi \; , \pi = - G \bar{\psi}i\gamma_5 \psi, \label{11}
\end{equation}
and the Lagrangian density (\ref{10}) reproduces Eq. (\ref{9}) upon
application of the constraints  (\ref{11}).
The effective action for the composite fields $\sigma $ and $\pi$ is
obtained by integrating over fermions in the path integral:
\begin{equation}
\Gamma(\sigma , \pi) = -i \mbox{Tr Ln } [i\gamma^{\mu}\partial_{\mu} -
(\sigma + i \gamma_5 \pi) ]
- \frac{1}{2G} \int d^D x (\sigma^2 + \pi^2). \label{12}
\end{equation}
The low energy dynamics are described by the path integral
(with the integrand $\exp(i\Gamma) $ ) over the fields $\sigma$ and $\pi$.
As $N_c \to \infty$, the path integral is dominated by the stationary points
of the action: $\frac{\delta \Gamma}{\delta \sigma} =
\frac{\delta \Gamma}{\delta \pi} = 0.$

Let us look at the effective potential in this theory. It is \cite{7}
\begin{equation}
V(\sigma , \pi) = \frac{4 N_c \Lambda^D}{(4\pi)^{D/2} \Gamma(D/2)}
\left[ ( \frac{1}{g} - \frac{1}{g_{cr}}) \frac{\rho^2}{2\Lambda^2}
+ \frac{2}{4-D} \frac{\xi_D}{D} (\frac{\rho}{\Lambda})^D \right]
+ O(\frac{\rho^4}{\Lambda^4}) , \label{13}
\end{equation}
where $\rho = (\sigma^2 + \pi^2 )^{1/2}, \xi_D = B(D/2-1,3-D/2),$
the dimensionless coupling constant $g$ is
\begin{equation}
g = \frac{4 N_c \Lambda^{D-2}}{(4\pi)^{D/2}\Gamma(D/2)} G, \label{14}
\end{equation}
and the critical coupling $g_{cr} = \frac{D}{2}-1$.

At $D>2$, one finds that
\begin{equation}
M^{(2)} \equiv \frac{d^2 V}{d \rho^2} \left. \right|_{\rho=0} \simeq
\frac{4 N_c \Lambda^{D-2}}{(4\pi)^{D/2}\Gamma(D/2)} \frac{g_{cr}-g}{g_{cr}g}.
\label{15}
\end{equation}
The sign of $M^{(2)}$ defines two different phases:
$M^{(2)} > 0$ $(g < g_{cr})$ corresponds to the symmetric phase and
$M^{(2)} < 0$ $(g > g_{cr})$ corresponds to the phase with spontaneous chiral
symmetry breaking, $U(1)_L \times U(1)_R \to U(1)_{L+R}$.
The value $M^{(2)} =0$ defines the critical point $g=g_{cr}$.

Therefore at $D>2$, a $\sigma $-model-like phase transition is realized.
However the case $D=2$ is special:
now $g_{cr} \to 0$ and $\xi_D \to \infty$ as $D \to 2.$ In this case
the effective potential is the well-known potential
of the Gross-Neveu model \cite{8}:
\begin{equation}
V(\sigma , \pi) = \frac{N_c}{2\pi g} \rho^2 - \frac{N_c \rho^2 }{2\pi}
 \left[ \ln \frac{\Lambda^2}{\rho^2} + 1 \right]. \label{16}
\end{equation}
The parameter $M^{(2)}$ is now :
\begin{equation}
M^{(2)} = \frac{d^2 V}{d \rho^2} \left. \right|_{\rho=0} \to + \infty.
\end{equation}
Therefore, in this model, one cannot use $M^{(2)}$ as a parameter governing
the continuous phase transition at $g = g_{cr} =0$ :
the phase transition is not a $\sigma$-model like phase transition
in this case.
Indeed, as follows from Eq. (\ref{16}),
the order parameter, which is a solution to the gap equation
$\frac{d V }{d \rho} =0$, is
\begin{equation}
\bar{\rho} = \Lambda \exp ( -\frac{1}{2g}). \label{18}
\end{equation}
in this model. The function $f(z)$, defined in Eq. (\ref{2}), is now
$f(g) = \exp (- \frac{1}{2g})$, i.e., $z=g$, and therefore the CPT
takes place in this model at $g=0$: $f(g)$ goes to zero only if
$g \to 0$ from the side of the non-symmetric phase.

Let us discuss this point in more detail.

At $D \geq 2$, the spectrum of the $\sigma$ and $\pi$ excitations
in the symmetric solution, with $\bar{\rho}=0$, is defined by the following
equation (in leading order in $\frac{1}{N_c}$) \cite{7}:
\begin{equation}
(\frac{1}{g} - \frac{1}{g_{cr}}) \Lambda^{D-2} + \frac{\xi_D}{2-D/2}
(-M_{\pi}^2)^{D/2-1} = 0. \label{19}
\end{equation}
Therefore at $D>2$, there are tachyons with
\begin{equation}
M_{\pi}^2=M_{\sigma}^2 = M^2_{tch} =
-\Lambda^2  (\frac{4 -D}{2\xi_D})^{\frac{2}{D-2}}
(\frac{g-g_{cr}}{g_{cr}g})^{\frac{2}{D-2}} \label{20}
\end{equation}
at $g > g_{cr}$, and  at $g < g_{cr}$ there are ``resonances'' with
\begin{equation}
|M_{\pi}^2| = |M_{\sigma}^2| =
\Lambda^2 (\frac{4 -D}{2\xi_D})^{\frac{2}{D-2}}
(\frac{g_{cr}-g}{g_{cr}g})^{\frac{2}{D-2}} \; , \label{21}
\end{equation}
which agrees with Eq. (\ref{8}).
\footnote{
For our purposes, it is sufficient to calculate the absolute value of
$M_{\pi}^2$. Notice that, as follows from Eq. (\ref{19}), narrow resonances
occur near $D=4$: $\frac{\Gamma}{M_R} \simeq \pi \frac{4-D}{D-2} \quad
(M_{\pi} = M_R - i \frac{\Gamma}{2} )$ .
}
Eq. (\ref{21}) implies that the limit $D \to 2$ is special.
One finds from Eq. (\ref{19}) that at $D=2$
\begin{equation}
M_{\pi}^2=M_{\sigma}^2 = M^2_{tch} = - \Lambda^2 \exp ( - \frac{1}{g})
\label{22}
\end{equation}
at $g> 0$, and
\begin{equation}
|M_{\pi}^2| = |M_{\sigma}^2| =  \Lambda^2 \exp (  \frac{1}{|g|})
\label{23}
\end{equation}
at $g< 0$, i. e., in agreement with the main feature of the CPT,
there are no light resonances in the symmetric phase at $D=2$.

The effective potential (\ref{16}) can be rewritten as
\begin{equation}
V(\sigma , \pi) = \frac{N_c \rho^2}{2\pi}
 \left[ \ln \frac{\rho^2}{\bar{\rho}^2}- 1 \right]  \label{24}
\end{equation}
(with $\bar{\rho}$ given by Eq. (\ref{18})) in the non-symmetric phase.
That is, in this phase $V(\sigma , \pi)$ is finite in the continuum limit
$\Lambda \to \infty$ after the renormalization of the coupling constant,
\begin{equation}
g=\frac{1}{\ln \frac{\Lambda^2}{\bar{\rho}^2}} \label{25}
\end{equation}
(see Eq. (\ref{18})). But what is the form of the effective potential
in the continuum limit in the symmetric phase, with $g<0$ ?
As Eq. (\ref{16}) implies, it is infinite as $\Lambda \to \infty$ :
indeed at $g<0$, there is no way to cancel the logarithmic divergence in $V$.

It is unlike the case with $D> 2$ : in that case, using Eq. (\ref{15}),
the potential (\ref{13}) can be put in a $\sigma$-model-like form :
\begin{equation}
V(\sigma , \pi) = \frac{M^{(2)}}{2} \rho^2
+ \frac{8 N_c}{(4\pi)^{D/2} \Gamma(D/2)} \frac{\xi_D}{(4-D)D} \rho^D.
\label{26}
\end{equation}
However, since $M^{(2)} = \infty$ at $D=2$, the $\sigma$-model like form
for the potential is not available in the Gross-Neveu model.

What are physical reasons of such a peculiar behavior of the effective
potential at $D=2$ ?
Unlike the case with $D>2$, at $D=2$ the Lagrangian density
(\ref{9}) defines a conformal theory in the classical limit.
By using the conventional approach, one can derive the following
equation for the conformal anomaly in this model :
\begin{equation}
\partial^{\mu} D_{\mu} = \theta_{\mu}^{\mu} =
\frac{\pi}{2 N_c} \beta (g)  \left[ (\bar{\psi}\psi)^2 + (\bar{\psi} i
\gamma_5 \psi )^2 \right] \label{27}
\end{equation}
where $D_{\mu}$ is the dilatation current, $\theta_{\nu}^{\mu}$
is the energy-momentum tensor, and the $\beta$ function
$
\beta = \frac{\partial g}{\partial \ln \Lambda}
$.
It is $\beta (g) = -g^2 $ both in the non-symmetric and
symmetric phases.
While the non-symmetric phase corresponds to asymptotically free
dynamics, the symmetric phase (with $g < 0$)
defines infrared free dynamics :
as $\Lambda \to \infty$, we are led to a free theory of massless
fermions, which is of course conformal invariant.

On the other hand, in the non-symmetric phase the conformal symmetry
is broken, even as $\Lambda \to \infty$.
In particular, Eq. (\ref{24}) implies that
\begin{equation}
\langle 0 | \theta_{\mu}^{\mu} | 0 \rangle = 4V(\bar{\rho}) =
- \frac{2N_c}{\pi}{\bar{\rho}}^2 \ne 0
\label{28}
\end{equation}
in leading order in $\frac{1}{N_c}$ in that phase.

The physics underlying this difference between the two phases is clear :
while negative $g$ correspond to repulsive interactions between
fermions, attractive interactions at positive $g$ lead to the formation of
bound states, thus breaking the conformal symmetry.

Notice the following interesting point. As follows from Eq. (\ref{26}),
at $D > 2$ the conformal symmetry is broken by a relevant
(superrenormalized) mass operator:
its dynamical dimension is $d=2$ at all $ 2 \leq D \leq 4$.
On the other hand, at $D=2$ the symmetry is broken by a marginal
(renormalized) operator with the dynamical dimension $d=2$.
This point is reflected in that while at $D=2$ the expression for
the order parameter $\bar{\rho}$ has an essential singularity at the
critical point $g=g_{cr}=0$, at $D>2$, the singularity at
$g=g_{cr}$ in $\bar{\rho}$ is not essential :
as follows from Eq. (\ref{13}), the solution to the gap equation
$\frac{d V}{d \rho} = 0 $ is $ \bar{\rho} \sim \Lambda
(g-g_{cr})^{\frac{1}{D-2}}$
in that case. As is known, the essential singularity implies the absence of
of fine tuning for bare parameters. This is another reason
why the CPT is so interesting.

Thus the CPT, in accordance with its name, describes the two essentially
different realizations of the conformal symmetry in the symmetric and
non-symmetric phases.

If one adds a fermion mass term, $m^{(0)} \bar{\psi} \psi $,
in the 2-dimensional GN model, the conformal and chiral symmetries will
be of course broken in both phases.
However, there remains an essential trace of the CPT also in this case :
an abrupt change of the spectrum of light excitations still takes place.
While now in the subcritical $(g< g_{cr}=0)$ phase repulsive interactions
between massive fermions
take place (and there are no light resonances there), in the supercritical
$(g > g_{cr} =0)$ phase the PCAC dynamics, describing interactions between
fermions and light $\pi$ and $\sigma$ bosons, is realized.
\footnote{
We are of course aware that the exact solution in the non-symmetric phase
of the 2-dimensional GN model yields a realization of
the Berezinsky-Kosterlitz-Thouless(BKT) phase :
though chiral symmetry is unbroken, the parameter $\bar{\rho}$ still
defines the fermion dynamical mass, and the would-be NG boson $\pi$
transforms into a BKT gapless excitation \cite{9}.
}

Besides the point that in the 2-dimensional GN model both
subcritical and supercritical phases are physical, this picture
is similar to that in QCD. It is hardly surprising: in both models
the dynamics in the supercritical phases are
asymptotically free. We will however
argue that the main features
of the CPT found in the GN model will retain valid (with
appropriate minor modifications) in the general case.

\section{The CPT in quenched QED4}

Another interesting example of the CPT is realized in
quenched QED4.
The dynamics in this model is relevant for some
scenarios of dynamical electroweak symmetry breaking and has been
intensively discussed in the literature (for a review see Ref. \cite{10}).
In the present consideration the emphasis of the discussion will be on the
points relevant for the general CPT in gauge theories.

We shall consider the ladder (rainbow) approximation in massless
QED4.  Since the contribution of fermion loops is omitted, the
perturbative $\beta$-function equals zero in this approximation.
However, as is well known \cite{10,11,12}, beyond the critical value
$\alpha = \alpha_c \sim 1$, there are nonperturbative divergences
which break the conformal symmetry in the model.  Moreover, since at
$\alpha = \alpha_c$, the anomalous dimension $\gamma_m$ of the chiral
operators $\bar{\psi} \psi$ and $\bar{\psi}  i \gamma_5 \psi$ is
$\gamma_m = 1$ \cite{12,13}, the four-fermion operators $(\bar{\psi} \psi)^2$
and $\bar\psi i \gamma_5 \psi)^2$ become (marginally) relevant:
their dynamical dimension $d$ is $d=
d_{c} - 2 \gamma_m = 4$, where
$d_{c} = 6$ is their canonical dimension.

Therefore, it is appropriate to include these four-fermion
operators in the QED action.  This leads to the gauged
Nambu-Jona-Lasinio model \cite{13}:
\begin{equation}
{\cal L} = - \frac{1}{4} \left(F_{\mu \nu} \right)^2 + \frac{1}{2}
\left[ \bar\psi, (i
\gamma^\mu D_\mu)\psi \right]
+ \frac{G}{2} \left[ (\bar\psi \psi)^2 + (\bar\psi i \gamma_5 \psi)^2
\right], \label{29}
\end{equation}
\noindent where $D_\mu = \partial_\mu - ie A_\mu$ (for simplicity, we consider
the chiral symmetry $U_L(1) \times U_R(1)$).  In this model, the gauge
interactions are treated in the ladder approximation and the
four-fermion interactions are treated in the Hartree-Fock (mean
field) approximation.

Since the coupling constant $G$ is dimensional, one may think that
the four-fermion interactions in Eq. (\ref{29}) explicitly break the
conformal symmetry.  The real situation is however more subtle.
The critical line in this model,
dividing the symmetric
phase, with the unbroken $U_L(1) \times U_R(1)$, and the phase
with the spontaneously broken chiral symmetry $(U_L(1) \times U_R(1)
\rightarrow U_{L+R}(1)$), was defined in Ref. \cite{14}.  Each point of the
critical line corresponds to a continuous phase transition.  We
distinguish two parts of the critical line:
\begin{equation}
g \equiv \frac{G\Lambda^2}{4 \pi^2} = \frac{1}{4} \left[ 1 + \left( 1 -
\frac{\alpha}{\alpha_c} \right)^{1/2} \right]^2, \: \:
\alpha_c = \frac{\pi}{3}, \label{30}
\end{equation}
\noindent at $g > \frac{1}{4}$, and
\begin{equation}
\alpha = \alpha_c \label{31}
\end{equation}
\noindent at $g < \frac{1}{4}$.  The anomalous dimension $\gamma_m$ of the
operators $\bar\psi \psi$ and $\bar\psi i \gamma_5 \psi$ along the
critical line is \cite{15}
\begin{equation}
\gamma_m = 1 + \left( 1 - \frac{\alpha}{\alpha_c} \right)^{1/2}. \label{32}
\end{equation}
\noindent In this approximation, the anomalous dimension of the
four-fermion operator $\left[ (\bar\psi \psi)^2 + (\bar\psi i
\gamma_5 \psi)^2 \right]$ equals $2 \gamma_m$.  Therefore
while this operator
indeed breaks the conformal symmetry along
the part (\ref{30}) of the critical line,
it is a marginal (scale
invariant) operator along the part of the critical line with $\alpha
= \alpha_c$:  its dynamical dimension is $d_{\bar\psi \psi} = 6 -
2\gamma_m = 4$ there.

Thus the part (\ref{31}) of the critical line with $\alpha = \alpha_c$ is
special.  In this case the symmetric phase is not only chiral
invariant but also conformal invariant.  On the other hand, in the
non-symmetric phase, both these symmetries are broken:  while the
chiral symmetry is broken spontaneously, the conformal symmetry is
broken explicitly  \cite{13,14}.

The effective action in this model is described in detail
in Refs. \cite{0,4}.
There is a similarity between the dynamics in quenched QED4 and
$D$-dimensional GN model considered in Sec. 3.
At $\alpha < \alpha_c = \frac{\pi}{3}$ $ (D>2)$ a $\sigma$-model-like
phase transition is realized in quenched QED4 (GN model);
at $\alpha = \alpha_c$ $(D=2)$ the CPT takes place in these models.
However, there is an essential difference between the CPT phase
transitions in these two models. While in the GN model,
the symmetric phase, with $g<0$, is infrared free, the symmetric phase in
quenched QED is a Coulomb phase, describing conformal invariant
interactions between massless
fermions and photons.

As was indicated in Sec. 3, a marginal operator is responsible for the
breakdown of the conformal symmetry in the non-symmetric phase in the
2-dimensional GN model (see Eq. (\ref{27})). This
leads to an essential singularity
in the expression for the order parameter $\bar{\rho}$ (\ref{18}).
This in turn cures fine tuning problem which takes place at $D>2$,
where relevant (superrenormalized) operators break the conformal symmetry.

A similar situation takes place in quenched QED4. While at
$\alpha < \alpha_c$, the (relevant) mass operator breaks the conformal
symmetry, at $\alpha = \alpha_c$, it is
broken (in non-symmetric) phase by a marginal operator \cite {0}.
In the next section, we shall summarize the main features of the
CPT.
We shall also discuss the phase transition in QED3.

\section{General Features of the CPT. A pseudo-CPT in QED3. }

Now we are ready to summarize the main features of the CPT.

There is an abrupt change of the spectrum of light excitations,
as the critical point $z = z_c$ is crossed, in the CPT.
As was shown in Sec. 2, this property is general and reflects the presence
of an essential singularity at $z=z_c$ in the scaling function $f(z)$.
This point is connected with the properties of $\beta$ function at
$z=z_c$. In the GN model, while $g_c=0$ is an ultraviolet stable fixed
point as $g\to g_c+0$ from the side of non--symmetric phase, it is an
infrared stable fixed point as $g\to g_c-0$ from the side of the symmetric
phase. In quenched QED4, the $\beta$ function
$\beta(\alpha)=-\frac{2}{3}\left(\frac{\alpha}{\alpha_c}-1\right)^{3/2}$
has a singularity at the critical point \cite{0}. We believe that these
two possibilities are typical for the CPT in general. (Recall that
the critical point $z=z_c$is an
ultraviolet stable fixed point in both
symmetric and non-symmetric phases in the case of a $\sigma$-model-like
phase transition).

The CPT is (though continuous) a non-$\sigma$-model-like phase transition.
This implies a specific form of the effective action, in particular,
the effective potential, for the light excitation near $z=z_c$.
While the potential does not exist in the continuum limit in the symmetric
phase, it has infrared singularities at $\rho = 0$ in the non-symmetric
phase ($\rho $ is a generic notation for fields describing the light
excitations).
As a result, unlike the $\sigma$-model-like
phase transition, one cannot
introduce parameters $M^{(2n)} = \frac{d^{2n} V}{d \rho^{2n}}
\left. \right|_{\rho = 0}$ which would govern the phase transition:
all of them are equal either to zero or to infinity.

The infrared singularities in the effective potential imply the presence
of long range interactions.
This is turn connected with
an important role of the conformal symmetry in the CPT.
In the examples considered in Sec. 3 and 4, while the symmetric phase is
conformal invariant, there is a conformal anomaly in the non-symmetric phase:
the conformal symmetry is broken by a marginal operator.
The latter allows to get rid of the fine tuning problem in such a dynamics
and provides a rich spectrum of light excitations in the
non-symmetric phase \cite{-1}.
We shall return to the problem of the effective action in the CPT
in the next section.

Because of the abrupt changing the spectrum of light excitations at
$z=z_c$, the very notion of the universality class for the dynamics
with the CPT seems rather delicate.
For example, in both GN model and QCD, at the critical point
$(g=0$ and $\alpha^{(0)}=0$, respectively), and at finite cutoff
$\Lambda$, the theories are free and their infrared dynamics
are very different from the infrared dynamics
in the non-symmetric phases of these theories (at $g>0$ and
$\alpha^{(0)}>0$, respectively).
This is a common feature of the CPT:
around the critical point, the infrared dynamics in the symmetric
and non-symmetric phases are very different. However, in the
non-symmetric phase,
the hypothesis of universality has to be applied to the region of
momenta $p$ satisfying $\bar{\rho} << p << \Lambda$, where $\bar{\rho}$
is an order parameter. In that region, critical indices
(anomalous dimensions) of both elementary and composite local operators
in near-critical regions of symmetric and non-symmetric phases are nearly
the same:
the critical indices are continuous functions of $z$ around $z=z_c$
\footnote{
However, because of explicit conformal symmetry breaking in
the non-symmetric phase, there are additional logarithmic factors
(such as $(\ln \frac{p}{\bar{\rho}})^c$) in Green's functions in that phase.
}.
On the other hand, since the infrared
dynamics (with $p \sim \bar{\rho}$ and
$p << \bar{\rho}$) abruptly changes as the critical
point $z=z_c$ is crossed, {\em the low energy effective actions in
the symmetric and non-symmetric phases are different}.

One can consider deformations of theories with the CPT, by adding relevant
operators in their Lagrangians, such as fermion mass terms,
which break explicitly the conformal symmetry. Also if there is a
perturbative running of the coupling in the symmetric phase, it will
lead to perturbative violation of the conformal symmetry.
In many cases, the deformations do not change the most characteristic
point of the CPT:
the abrupt change of the spectrum of light excitations at
$z=z_c$ discussed above. The reason is that there is an additional,
nonperturbative, source  of the breakdown of the conformal symmetry
in the non-symmetric phase, which provides the creation of light
composites.

The conception of the CPT, in a slightly modified form, can be also
useful for a different type of dynamics. As an example, let us consider
QED3 with massless four-component fermions \cite{20}.
It is a superrenormalizable theory where ultraviolet dynamics plays
rather a minor role. As was shown in Refs. \cite{21,22}, when the number of
fermion flavors  $N_f$ is less than $N_{cr}$, with $3< N_{cr} < 4$,
there is dynamical breakdown of the flavor $U(2N_f)$ symmetry in the model,
and fermions acquire a dynamical mass
\footnote{
We are aware that there is still a controversy concerning this result:
some authors argue that the generation of a fermion mass occurs at all
values of $N_f$ \cite{23}. For a recent discussion supporting the
relation (\ref{85}), see Ref. \cite{24}.
}:
\begin{equation}
m_{dyn} \sim \alpha_3 \exp \left[ -\frac{2\pi}{\sqrt{N_{cr} / N_f
-1}} \right],
  \label{85}
\end{equation}
where the coupling constant
$\alpha_3 = e^2/4\pi$ is dimensional in QED3.

Though this expression resembles the expression for
the dynamical mass in quenched QED4 \cite{12}, where $\Lambda$ plays the role
of
$\alpha_3$ and $\alpha$ plays the role of $N_f$, the phase transition at
$N_f=N_{cr}$ is, strictly speaking, not the CPT.
Indeed, because of superrenormalizability of QED3, the ultraviolet cutoff
$\Lambda$ is irrelevant for the dynamics leading to relation (\ref{85}).
Also, since $\alpha_3$ is dimensional, the conformal symmetry is broken
in both symmetric $(N_f > N_{cr})$ and non-symmetric
$(N_f < N_{cr})$ phases.

Nevertheless, the consideration of the spectrum of light
(with $M^2 << \alpha_3^2$)
excitations in this model can be done along the
lines used in Sec. 2.
In agreement with the result of Ref. \cite{6}, where the BS equation
was used, one concludes that there are no light resonances
(with $M^2 << \alpha_3^2$ ) in the symmetric phase of QED3 and
that there is an abrupt change of the spectrum of light
excitations at $N_f = N_{cr}$.

It is appropriate to call the phase transition in QED3 a
pseudo-CPT: in the non-symmetric phase, at $N_f < N_{cr}$, a new,
nonperturbative, source of the breakdown of the conformal symmetry
occurs.

\section{The Effective Action in Theories with
the CPT and the Dynamics of the Partially Conserved Dilatation
Current}

In this section we shall discuss the properties of the effective
action in theories with the CPT in more detail.
In particular we shall consider a connection of the dynamics of the CPT
with the hypothesis of the partially conserved dilatation current
(PCDC) \cite{25,26,27,41}.

The effective potentials derived in the 2-dimensional GN model
(see Eqs. (\ref{16})) and (\ref{24})) and in quenched QED4
with $(\alpha , g) = (\alpha_c, \frac{1}{4})$
\cite{0} have a similar form.

Moreover, one can show that the kinetic term and terms with
higher number of derivatives in both the GN model and quenched QED4
are conformal invariant \cite{0,8}.
In other words, the conformal anomaly comes only from the effective
potential in both these models.

This point is intimately connected with the PCDC dynamics.
In order to see this, let us determine the divergence of the dilatation current
in these models. Eq. (\ref{24}) implies that
\begin{equation}
\partial^{\mu} D_{\mu} = \theta^{\mu}_{\mu} =
-\frac{2N_c}{\pi} \rho^2 \label{86}
\end{equation}
in the GN model, and
\begin{equation}
\partial^{\mu} D_{\mu} = \theta^{\mu}_{\mu} =
- \frac{\tilde{A}^2}{4 \pi^2} m_{dyn}^2 \rho^2   \label{87}
\end{equation}
in quenched QED4 with $\alpha = \alpha_c$, where the CPT takes place
$(m_{dyn} \equiv \bar{\Sigma}_0)$ \cite{0}.
Now, recall that the dynamical dimension $d_{\rho}$ of the field $\rho$ is
$d_{\rho}=1$ and $d_{\rho}=2$ in the GN model and in quenched QED4 (with
$\alpha = \alpha_c$), respectively.
Therefore Eqs. (\ref{86}) and (\ref{87}) assure that the dynamical
dimension of the operator $\theta^{\mu}_{\mu}$ coincides with its
canonical dimension:
$d_{\theta}=2$ and $d_{\theta}=4$ in the 2-dimensional GN model
and quenched
QED4, respectively.
This implies the realization of the PCDC hypothethis in these
models  \cite{25,26,27,41}: the operator $\theta^{\mu}_{\mu}$
has the correct transformation properties under dilatation
transformations.

In the renormalization group language, this means that the conformal
symmetry in these models is broken by marginal (renormalized) operators
and not by relevant (superrenormalized) ones (irrelevant
(nonrenormalized) operators contribute only
small corrections in the infrared
dynamics).

Though these two models are very special, one may expect that at least
some features of this picture will survive in the general case of theories
with the CPT. In particular, one may expect that in the general case
the effective potential has the form
\begin{equation}
V(\rho ) = C \bar{\rho}^D \left(
\frac{\rho}{\bar{\rho}} \right)^{\frac{D}{d_{\rho}}} F(
\ln \frac{\rho}{\bar{\rho}} ) \label{88}
\end{equation}
where $C$ is a dimensionless constant and $F(x)$ is a (presumably)
smooth function.

The contribution of $V(\rho)$ (\ref{88}) into the conformal anomaly
is of the form
\begin{equation}
\theta^{\mu}_{\mu} \sim \bar{\rho}^D \left( \frac{\rho}{\bar{\rho}}
\right)^{\frac{D}{d_{\rho}}}
F'(\ln \frac{\rho}{\bar{\rho}}),   \label{89}
\end{equation}
where $F'(x) = \frac{d F}{d x}$, i.e., in the general case, logarithmic
factors may destroy the covariance (with respect to dilatation
transformations) of the relation for the conformal anomaly \cite{0}.

Also, one should expect that the conformal invariance of the kinetic term
and terms with higher number of derivatives may
also be destroyed by logarithmic terms.

It is clear that the effective action in theories with the CPT
are very different from that in the 4-dimensional
linear $\sigma$-model and
Nambu-Jona-Lasinio model, where the conformal symmetry is broken
by relevant operators and the chiral phase transition is a mean-field one.

This point can be relevant for the description of the low energy dynamics
in QCD and in models of dynamical electroweak symmetry breaking.
In particular, as was already pointed out in Ref. \cite{27}, the
low energy dynamics are very sensitive to the value of the dynamical
dimension $d_{\rho}$.

\section{Conclusion}

In this talk I discussed the conception of the conformal phase
transition (CPT) which provides a useful framework for studying
nonperturbative dynamics in gauge (and also other) field theories.
We described the general features of this phase transition.

The CPT is intimately connected with the nonperturbative breakdown
of the conformal symmetry, in particular, with the PCDC dynamics.
In the non-symmetric phase the conformal symmetry is broken by
marginal operators. This in turn yields a constraint on the form
of the effective action in theories with the CPT.

In all the examples of the CPT considered in this paper, the
conformal symmetry was explicitly broken by the conformal anomaly
in the phase with spontaneous chiral symmetry breaking. Is it
possible to realize dynamics with both chiral and conformal
symmetries being broken spontaneously? Although at present this
question is still open, we would like to note that long ago
arguments had been given against the realization of such a
possibility \cite{42}.

The conception of the CPT can be useful for strong-coupling gauge theories,
in particular, for QCD and models of dynamical electroweak symmetry
breaking.
In connection with that, we note that the effective action considered
in Sec. 6 may be relevant for the description of $\sigma$ meson
$( f^0 (400 - 1200))$ \cite{36,37}. If it is rather light
(with $M_{\sigma} \simeq 600$ MeV) as some authors conclude \cite{37},
it can dominate in the matrix elements of the operator
$\theta_{\mu}^{\mu}$ in low energy dynamics, i. e., it can be considered
as a massive dilaton, as was already suggested some time
ago \cite{25,27}.

It is also clear that the conception of PCDC and massive dilaton can
be useful for the description of the dynamics of composite Higgs boson.

As it is discussed in detail in Ref. \cite{0}, a very interesting
example of the CPT may be provided by the phase transition with
respect to the number of fermion flavors in a $SU(N_c)$ vector-like
gauge theory in 3+1 dimensions \cite{2,5}.

Another application of the CPT (or pseudo-CPT)
may be connected with non-perturbative dynamics
in condensed matter. Here we only mention the dynamics of non-fermi
liquid which might be relevant for high-temperature surperconductivity:
some authors have suggested that QED3 may serve as an effective theory of
such a dynamics \cite{39}.

There has been recently a breakthrough in understanding non-perturbative
infrared dynamics in supersymmetric (SUSY) theories
(for a review see Ref. \cite{40} ). It would
be worth considering the realization of the CPT, if any, in SUSY theories,
thus possibly establishing a connection between SUSY and non-SUSY dynamics.

\end{document}